# When Algorithms Infer Gender: Revisiting Computational Phenotyping with Electronic Health Records Data

## Authors

Jessica Gronsbell[1,*], Hilary Thurston[2], Lillian Dong[3], Vanessa Ferguson[4], Diksha Sen Chaudhury[1], Braden O'Neill[5], Katrina S. Sha[1], Rebecca Bonneville[6]

[1] Department of Statistical Sciences, University of Toronto, Toronto, M5G 1X6, ON, Canada
[2] Department of Gender, Feminist, and Women's Studies, York University, Toronto, M3J 1P3, Ontario, Canada
[3] Department of Biostatistics, Dalla Lana School of Public Health, University of Toronto, Toronto, ON, M5S 1A8, Canada
[4] School of Health Policy & Management, York University, Toronto, ON, M3N 3A7, Ontario, Canada
[5] Department of Family Medicine, University of British Columbia, Vancouver, BC, V6T 1Z3
[6] Resilient Minds Psychiatry, Ojai, 93023, California, U.S.

* Correspondence to [j.gronsbell@utoronto.ca](j.gronsbell@utoronto.ca)

# Abstract

Computational phenotyping has emerged as a practical solution to the incomplete collection of data on gender in electronic health records (EHRs). This approach relies on algorithms to infer a patient's gender using the available data in their health record, such as diagnosis codes, medication histories, and information in clinical notes. Although intended to improve the visibility of trans and gender-expansive populations in EHR-based biomedical research, computational phenotyping raises significant methodological and ethical concerns related to the potential misuse of algorithm outputs. In this paper, we review current practices for computational phenotyping of gender and examine its challenges through a critical lens. We also highlight existing recommendations for biomedical researchers and propose priorities for future work in this domain.

# Highlights

- Sex and gender are inconsistently recorded in electronic health records (EHRs), limiting the scope of biomedical research using these data.
- Computational phenotyping algorithms attempt to fill these gaps by inferring gender-related information from patients' historical health data.
- While these approaches aim to improve the visibility of trans and gender expansive people in biomedical research, they also introduce important methodological and ethical concerns, including (1) data quality issues, (2) underlying assumptions about gender, (3) bias in algorithm design and validation, and (4) potential for misuse.
- Future research should focus on building just and conceptually sound foundations for gender-based inquiry, such as creating and using measurement tools that accommodate fluidity, center lived experience rather than biological proxies, and allow for individualized data collection without defaulting to gender assignment.

# Keywords

# Background

*"The light of big data creates big shadows."* [1]

Data from electronic health records (EHRs) are foundational to biomedical research, underpinning studies across clinical medicine, public health, genomics, and health services research [2–6]. Sex and gender are widely recognized as critical variables for understanding health and illness and are often mandated for collection in EHRs [7–13]. EHR systems can capture sex and gender information across multiple fields with distinct clinical and operational meanings, including gender identity, sex assigned at birth, and legal or administrative sex. However, these fields, if available at all, are inconsistently populated across healthcare settings [14–20]. A study of 1.5 million adult patients at Mass General Brigham in the U.S. found that while legal sex was recorded for all patients, only 20% had information on gender identity or sex assigned at birth [21]. Similarly, in Ontario, Canada, just 0.8% of nearly 400,000 adult primary care patients had gender identity documented in their records [22]. At Rush University Medical Center in the U.S., only 25% of nearly 50,000 unplanned hospital admissions records included a populated gender identity field [16].

In response to these gaps, *computational phenotyping* has emerged as a means to augment incomplete data collection, particularly for gender-related information [23]. A computational phenotype is an algorithm that infers a patient's gender based on information in their health record, such as diagnosis codes, medication codes for hormone prescriptions, procedure codes for gender-affirming care, and keywords in clinical notes [24–35]. This approach is a pragmatic solution to the limited uptake of sex and gender fields and has been used to collect data on trans and gender-expansive populations who have been historically underrepresented in EHR-based studies and biomedical research more broadly [36]. However, computational phenotypes bring significant methodological and ethical challenges that call into question both their validity and utility [37–40], including issues of data quality, embedded assumptions about gender, bias in algorithm design and validation, and risks of misuse.

As a diverse, interdisciplinary group of researchers, clinicians, and scholars, we occupy social locations that situate us at intersections of gender identity and expression, sexual orientation, race, and class. Together, we draw on experiential knowledge that reflects both oppression and privilege, alongside our professional and academic expertise across multiple theoretical and conceptual frameworks. These combined perspectives inform our critical evaluation of the methodological and ethical dimensions of computational phenotyping of gender. We ground our analysis in broader sociopolitical perspectives to emphasize the importance of considering the contexts in which computational phenotypes are developed and applied. We also outline recommendations for biomedical researchers and identify priorities for future research.

# Review of current practices

## Overview of computational phenotyping

EHR-based biomedical studies rely on phenotyping, the process of identifying patients with particular characteristics or conditions (i.e., phenotypes) using data in their health records [41,42]. Phenotypes are used to identify study populations as well as to extract variables for analysis [43]. When phenotype information is not explicitly available in a structured field (e.g., age or a lab test result), researchers develop a computational phenotyping algorithm to infer the phenotype based on a combination of structured and unstructured EHR data, such as diagnosis codes, medication prescriptions, and information in clinical notes (see **Figure 1**).

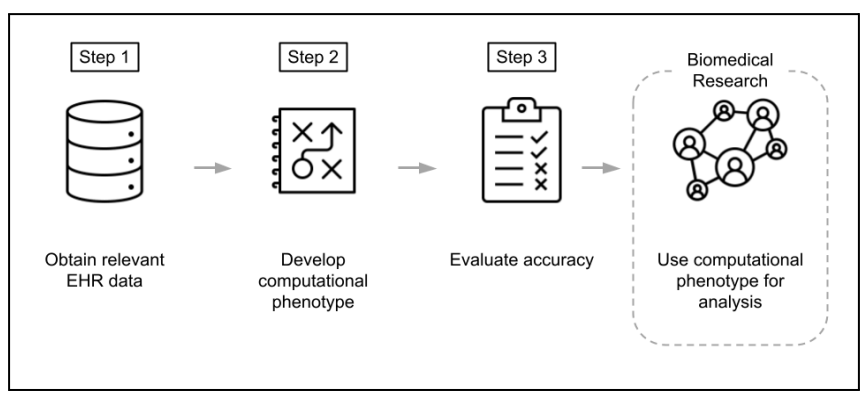

**Figure 1**: Overview of the computational phenotyping process: First, relevant structured (e.g., medical codes) and unstructured (e.g., clinical notes) data are obtained. Next, a rule-based or machine learning algorithm is used to develop the computational phenotype. Lastly, its accuracy is evaluated against a gold-standard label. The phenotype is then used in biomedical research to identify a study population or to derive an analytic variable.

Computational phenotypes can be developed through rule-based algorithms based on expert-derived criteria (e.g., a patient has the phenotype if their record contains relevant diagnosis and medication codes) or through machine learning models trained to predict phenotypes based on patterns in EHR data [44]. In either case, researchers validate the accuracy of the computational phenotype against a gold-standard label to ensure that it is suitable for downstream research. Gold standards are typically obtained through manual review of patient records, but may also be derived from lab test results or patient-reported data depending on the phenotype [45]. Accuracy is evaluated using standard performance metrics, including true and false positive rates and positive and negative predictive values [46–48]. Provided the computational phenotype is sufficiently accurate, it is then used as the basis for biomedical research. That is, it may be used to identify a study population (e.g., patients classified as having asthma by the algorithm for a study of treatment efficacy) and/or to derive variables of interest for analysis (e.g., a binary indicator of asthma status in a study of respiratory diseases).

# Computational phenotypes for gender

## Motivation

While computational phenotyping has traditionally focused on common, chronic illnesses (e.g., asthma, heart failure), it has more recently been used to infer non-clinical characteristics due to incomplete or inconsistent documentation [43,44,49]. In the context of gender, existing algorithms primarily aim to identify trans[1] and other gender expansive individuals in order to identify study populations for biomedical research studies [24,51]. Historically, data on these populations is extremely limited due to structural oppression, including harassment and harm related to disclosure of identity, sparse research funding, and barriers faced by transgender researchers [52,53]. At the same time, existing studies show that transgender populations have disproportionately high rates of mental health distress, substance use, and HIV relative to cisgender populations [51]. Recognizing these disparities, several federal agencies have issued calls for action. A 2011 report from the Institute of Medicine (now the National Academy of Medicine) emphasized the need for more research at the intersection of LGBTQ+ health and racial/ethnic minority health [54], and in 2016, the director of the National Institute on Minority Health and Health Disparities designated gender minorities as a "health disparity population for research purposes" [55]. EHRs, with their rich longitudinal and real-world data, offer a unique opportunity to study these populations at a scale and level of depth not possible in earlier research [28].

## Existing algorithms

We identified 12 studies proposing gender computational phenotypes based on a previous narrative review [23]. With the exception of the study by Hua et al., existing algorithms are rule-based, relying on combinations of medical codes, keywords, and/or sex and gender fields (see **Table 1**).  For example, Roblin et al. identified transgender individuals at Kaiser Permanente Georgia using an algorithm derived from fields such as diagnosis codes (e.g., codes related to sexual and gender disorders) and gender specific keywords in clinical notes (e.g., "transgender", "transsexual", "gender dysphoria") [24].  Further information such as procedure codes was then used to discern female to male (FTM) and male to female (MTF) identity (e.g., codes for hysterectomy) [24]. Ehrenfeld et al. employed a similar approach within Vanderbilt University Medical System, identifying transgender people on the basis of having at least one relevant diagnosis code or keyword selected from previous literature, the authors' expertise, and billing practices at the time [25]. Foer et al. introduced several computational phenotypes based on diagnosis codes, keywords, gender identity fields, and discrepancies between gender identity, legal sex, and sex assigned at birth fields at Partners Healthcare (now Mass General Brigham) in Boston, Massachusetts [26].  Xie et al. similarly utilized diagnosis codes and keywords to identify individuals as "definitely", "probably", or "not" transgender using data from Kaiser Permanente Southern California [28].

---

[1]See Table 4 for definitions of gender modalities. Occasionally transgender is shortened to trans [50].

| Study | Data Source | Gold Standard Label | Rule-based Algorithm | Data Used for Algorithm Development | | | | |
|---|---|---|---|---|---|---|---|---|
| | | | | Diagnosis codes | Medication Codes | Procedure Codes | Gender & Sex Fields | Clinical Notes |
| Roblin et al. (2016) | Kaiser Permanente Georgia | TG (MTF, FTM) | ✔ | ✔ | | ✔ | | ✔ |
| Ehrenfeld et al. (2019) | Vanderbilt University Medical Center | TG | ✔ | ✔ | | | | ✔ |
| Foer et al. (2020) | Partners Healthcare | TG | ✔ | ✔ | | | ✔ | ✔ |
| Chyten-Brennan et al. (2020) | Montefiore Health System | TGNB | ✔ | ✔ | ✔ | | ✔ | ✔ |
| Xie et al. (2020) | Kaiser Permanente Southern California | TG | ✔ | ✔ | | | | ✔ |
| Alpert et al. (2021) | CancerLinQ | TGNB | ✔ | ✔ | | | ✔ | |
| Guo et al. (2021) | University of Florida Health | TGNC (TM, TF, unknown) | ✔ | ✔ | ✔ | ✔ | ✔ | ✔ |
| Wolfe et al. (2021) | Veterans Health Administration | TG | ✔ | ✔ | ✔ | | ✔ | |
| Streed et al. (2023) | Fenway Health | TGD (TGM, TGW) | ✔ | ✔ | ✔ | ✔ | | |
| Hua et al. (2023) | Mass General Brigham | TGD | | ✔ | | ✔ | ✔ | ✔ |
| Hines et al. (2023) | University of Iowa Hospitals and Clinics | GE | ✔ | ✔ | ✔ | | ✔ | |
| Kim et al. (2024) | Pediatric Emergency Department | TGNB | ✔ | | | | ✔ | ✔ |

**Table 1**: Overview of computational phenotyping algorithms for gender. Abbreviations: TG = Transgender, MTF = Male to Female, FTM = Female to Male, TGNB = Transgender and Nonbinary, TGNC = Transgender and Gender-Nonconforming, TM = Transmasculine, TF = Transfeminine, TGD = Transgender and Gender Diverse, TGM = Transgender Men, TGW = Transgender Women, GE = Gender Expansive.

More recently, Chyten-Brennan et al. developed an algorithm to identify transgender and non-binary patients from Ryan White-funded clinics that provide dedicated HIV care within Montefiore Health System, the largest healthcare system in the Bronx neighborhood of New York City [27]. The algorithm supplemented diagnosis codes and keywords with gender-affirming medication prescriptions (e.g., concurrent male gender marker and estrogen prescription) and gender variables systematically reported to receive Ryan White HIV/AIDS Program funding (e.g., yes/no field for "transgender") [27]. Within CancerLinQ, a database on people with cancer across practices within the U.S., Alpert et al. used diagnosis codes for gender identity disorder or transsexualism and variables derived from structured gender fields together with diagnosis codes (e.g., male gender and malignant neoplasm of the vulva) to identify transgender and nonbinary people [29].

Within academic medical centers, Guo et al. identified transgender and gender nonconforming individuals in the University of Florida Health Integrated Data Repository using a combination of diagnosis codes, keywords, medication prescriptions, demographic information, and procedure codes related to gender-affirming surgeries [30]. At the University of Iowa Hospitals and Clinics, Hines et al. developed an algorithm to identify gender expansive individuals, including those who identify as transgender, nonbinary, transgender male or female, and other identities [34]. Their approach relied on discrepancies between legal sex, sex assigned at birth, and gender identity (excluding missing fields), as well as diagnosis codes for gender dysphoria or unspecified endocrine disorders and medication codes for estradiol or testosterone, which may indicate gender-affirming care [34]. Wolfe et al. used an analogous approach based on diagnosis codes related to gender identity disorder and variables derived from codes for unspecified or not otherwise specified endocrine disorders, use of gender-affirming hormone therapy (i.e., hormones not associated with documented sex), and changes in the sex field within the US Department of Veterans Affairs Health System [31].

In contrast to the aforementioned studies that developed algorithms for entire populations of patients within a particular healthcare system, database, or institution, Streed et al. narrowed the scope of their study to evaluate the performance of a previously unvalidated algorithm for identifying transgender and gender diverse people with self-reported gender-identity data at Fenway Health in Boston, Massachusetts [32]. Similar to prior studies, the algorithm was based on the presence of transgender-related diagnosis and procedure codes as well as gender-affirming prescription data [32]. In another line of work, Kim et al. developed a computational phenotype to identify transgender and nonbinary individuals within a pediatric emergency department in the U.S. using keywords and gender and sex fields [35].

In recent years, machine learning methods have gained popularity as rule-based algorithms can be prohibitively resource-intensive to develop due to the complexity and variability of clinical documentation [44]. Only one study applied a machine learning approach to identify transgender and gender diverse patients within the Mass General Brigham healthcare system. Hua et al. first screened patients using sex and gender fields and medication prescriptions and then applied ClinicalBERT, a variation of bidirectional encoder representation from transformers (BERT) that has been pre-trained on biomedical text [33].

# Methodological and ethical issues

Despite the growing body of research proposing computational phenotypes, their development and application raise important methodological and ethical concerns. Much of the literature is thoughtful and self-critical, and we draw on many of the stated limitations within the reviewed studies to outline four central challenges: data quality issues, underlying assumptions about gender, bias in algorithm design and validation, and potential for misuse.

## Data quality

*"[B]efore there are data, there are people…" [56]*

Information about gender recorded in EHRs is generally incomplete and inaccurate [57–60]. It is shaped by a complex interplay of factors, including the types of care individuals seek or are able to access, what they disclose during clinical encounters, and how healthcare institutions and providers document and interpret that information (see **Figure 2**). As a result, the data do not reflect an individual's self-identified gender, but rather how that identity is filtered through institutional practices and systemic bias [57,61].

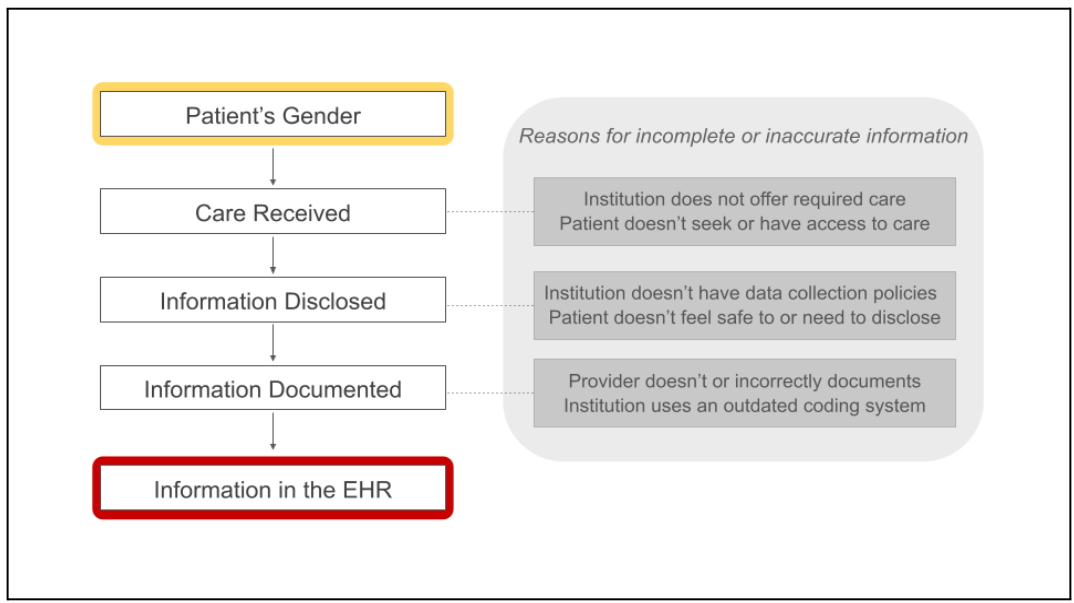

**Figure 2**: Process by which gender-related information enters an individual's EHR and a non-exhaustive list of reasons for incomplete or inaccurate information.

### Systemic and institutional-level factors

At the institutional level, the collection of sex and gender data, if implemented at all, is often executed without reference to best practices [16,19,62]. It is frequently confined to specific clinical contexts such as psychiatry, endocrinology, or gender clinics, and carried out without adequate provider training [14]. For example, in developing a computational phenotype for transgender and nonbinary individuals with cancer, Alpert et al. found that most oncologists either do not ask about gender identity or do so in ways that make patients uncomfortable responding [29]. Additionally, many of the data elements used for computational phenotyping, such as diagnosis and procedure codes, are derived from pathologizing and outdated documentation practices. For instance, "transgender" continues to appear on problem lists used to track current medical conditions and was a key component of an algorithm for identifying transgender patients within Partners Healthcare System [26]. Ehrenfeld et al. also discussed using diagnosis codes that contain outdated transgender-related terms, such as a code for "trans-sexualism with heterosexual history" [25]. These data do not represent individuals who do not pursue gender-affirming care, decline formal diagnosis, or who face structural barriers to accessing care, while also misrepresenting those who are documented in inaccurate or stigmatizing ways.

More generally, these issues illustrate the slow pace of institutional and coding reforms. It was not until 2019, with the release of the World Health Organization's 11th edition of the International Classification of Diseases (ICD-11), that widely used diagnosis codes such as F64 (Gender identity disorders) and F65.1 (Fetishistic transvestism) were replaced by HA60 (Gender incongruence of adolescence or adulthood) [63]. This revision moved gender incongruence out of the mental disorders chapter and into one focused on sexual health, reflecting new "knowledge that trans-related and gender diverse identities are not conditions of mental ill-health, and that classifying them as such can cause enormous stigma" [63]. However, adoption of ICD-11 has been uneven globally, partly due to the complexity of transitioning from ICD-10. For example, the United States has no firm timeline for ICD-11 implementation and took more than 20 years to complete the shift from ICD-9 to ICD-10 [64]. That said, even with widespread usage of more affirming coding standards, outdated codes will remain in patient records and continue to shape computational phenotypes.

Meanwhile, federal policies have changed what care institutions can offer, and in turn, what information is recorded in EHRs [65–67]. As part of a broader trend of banning gender-affirming care [68], the Veterans Health Administration has begun phasing out medical treatments for gender dysphoria in accordance with President Trump's "Defending Women from Gender Ideology Extremism and Restoring Biological Truth to the Federal Government" executive order [69]. Consequently, diagnosis codes that often underpin computational phenotyping algorithms to identify transgender people, such as those developed using Veterans Health Administration data in the work of Streed et al, will soon be erased [32].

Provider- and patient-level factors

At the provider level, documentation practices can reflect clinicians' assumptions and personal understanding rather than patients' self-identified gender. Ehrenfeld et al. noted that many providers misunderstood trans identities in their chart review process, finding ambiguous documentation of identities and pronouns used in non-affirming ways (e.g., "(s)he") [25]. Guo et al. identified a similar phenomenon and referenced a clinical note wherein a provider misunderstood trans female, stating that a patient was "a male who is trans female (born female living as male) and currently taking testosterone cypionate for male hormone" [30]. Misrepresentation or misunderstanding of a patient's gender, whether intentional or unintentional, reflects a manifestation of structural bias that can delegitimize patients' identities, contribute to clinical mistrust, and perpetuate inequities in care [70]. This bias also inevitably compromises the data for computational phenotyping by distorting the representation of gender in patient records and, in turn, leads to harmful misclassifications in algorithms' outputs.

At the patient level, trans people disproportionately experience mistreatment in healthcare settings, with 24% of respondents in the 2022 U.S. Trans Survey reporting avoidance of care due to fears of being mistreated and another 24% not disclosing their gender to their healthcare providers [71]. Common negative experiences cited by patients include bias, discrimination, and disparaging comments from healthcare providers [12,72,73]. These experiences are further amplified by intersecting systems of oppression, including racism, sexism, ageism, and classism, that not only impact the care patients receive, but what information they disclose and how that information is documented in their EHR [74].

For example, Chyten-Brennan et al. found that their algorithm for identifying transgender and nonbinary people within HIV/AIDS clinics within Montefiore Health System was significantly less accurate for Hispanic people [27]. The authors suggest that disparities in data capture, particularly for immigrant and non-English-speaking communities, reflect broader systemic barriers to equitable care and documentation. Moreover, limited engagement and access to healthcare, especially for those at intersecting forms of marginalization, further contributes to the incomplete and inaccurate capture of gender information in EHRs. Chyten-Brennan et al. also found that less than 1% of individuals were confirmed as transgender or nonbinary by their algorithm, which is significantly lower than anticipated [27]. The authors attribute this finding to stigma among HIV providers as well as disparate care engagement and disclosure among transgender and nonbinary people.

## Assumptions about gender

*"Not everyone is male or female. Not everyone is cis or trans. The sooner we make space for these truths, the better." [75]*

Trans people are those "whose gender or gender expression differs from expectations associated with the sex assigned to them at birth" [76]. As Os Keyes writes in *The Misgendering Machines*, this notion of difference encompasses a wide range of identities and experiences, including binary transitions, nonbinary or genderfluid identities, and people who don't identify with any gender [77]. While gender theorists hold differing views on the nature of gender, they broadly agree that it is not "immutable, binary, or intrinsically linked to physiology" [77]. These insights challenge the core, though often implicit, assumption in phenotyping studies that gender is a fixed and essential trait that can be reliably extracted from historical data in a patient's health records. This assumption is reflected in the use of oversimplified categorization schemes and in the ambiguous way gender is often operationalized in existing studies.

### Categorization of gender

Many computational phenotypes adopt a binary classification scheme of "transgender" or "transgender or nonbinary" versus "not" [25–27,29,31,33–35]. This model tacitly treats gender as static and singular, erasing its temporal and contextual variability and misrepresenting the lived experiences of many people. It also reinforces the false notion that gender must be stable to be measurable [78,79]. Streed et al. explicitly acknowledge that a fundamental limitation of their algorithm is its inability to accommodate changes in gender over time [32].

Seeking a more granular approach to gender classification, three studies further categorized individuals identified by their algorithm. Roblin et al. used a binary categorization of MTF and FTM [24] while Streed et al. classified individuals receiving hormone therapy as either transgender men or transgender women [24,32]. Guo et al. utilized three categories: transmasculine, transfeminine, and unknown [30]. While there have been considerable changes in terminology over the last several decades, with transmasculine and transfeminine becoming increasingly popular, categories aimed at identifying the "directionality of transness" can unnecessarily binarize nonbinary people and misportray those with additional genders (e.g., man, two-spirit) [14,37]. There is rich literature within sociology and informatics on best practices for categorizing and collecting data on gender [80–82]. The Williams Institute has developed two-step approaches for health surveys that first ask individuals if they identify within the binary and then follow with questions about transgender status [83]. Kronk et al. proposed a similar two-step collection process designed specifically for EHR systems, first inquiring about gender identity and then about the gender marker on an individual's birth certificate [14]. When used to supplant self-report data, which is often regarded as the most accurate source of truth within EHR-based research [20,84,85], computational phenotyping models should be held to similar standards. However, phenotyping algorithms must inevitably work backwards from administrative or clinical indicators, such as diagnosis and procedure codes, that often reflect medical intervention rather than identity itself. This backward approach not only increases the risk of misclassification, but also reinforces a medicalized framing of gender and diminishes individuals' agency in defining their own gender.

### Operationalization of gender

The limitations surrounding the categorization of gender highlight the broader question of what computational phenotypes aim to measure. We have been deliberate in using the general term "gender" throughout our discussion[2] as the output of phenotyping algorithms is often unclear and may reflect gender identity, expression, medical transition, or some combination thereof. Each of these concepts carries distinct implications for health and requires careful consideration when applied in biomedical research [86]. Gender identity refers to one's internal sense of self and how one identifies, while gender expression involves the outward presentation of gender through appearance and behavior. Computational phenotypes, both in their design and in their validation, rely on proxies for these concepts, which are inherently imperfect and often overlap.

---

[2] However, when referencing publishing articles, we adhere to their original terminology.

For example, clinical notes may inconsistently or inaccurately capture gender expression or identity, self-reported information in sex and gender fields may conflate identity with administrative categories, and procedure codes related to medical transition capture only certain interventions and do not fully represent a person's gender identity or experience. While many algorithms combine these data types to improve sensitivity, this approach comes at the cost of perpetuating an ambiguous operationalization of gender.

## Algorithm design and validation

*"Algorithms are opinions embedded in code." [87]*

The aforementioned challenges related to data quality, combined with assumptions about gender underlying computational phenotypes, introduce bias into both algorithm design and validation. These biases can foster overconfidence in an algorithm's outputs and lead to flawed conclusions in EHR-based studies that rely on computational phenotypes.

### Bias in design

Computational phenotypes are designed to capture individuals whose clinical encounters follow predictable and codifiable patterns. Many existing algorithms assume that gender can be inferred from clinical, biological, or administrative markers such as diagnosis codes, hormone prescriptions, and gender-affirming procedures. While this reliance on available EHR data is pragmatically necessary, it is also inherently reductive and pathologizing, as it encodes gender entirely within biomedical ontologies. For instance, diagnosis codes for gender dysphoria, transsexualism, or unspecified endocrine disorders often reflect reimbursement practices, medical necessity determinations, or outdated documentation standards. Many of these codes originate from historically pathologizing frameworks, including earlier versions of the ICD and Diagnostic and Statistical Manual of Mental Disorders (DSM), which classified gender diversity as a form of mental illness or sexual deviance [63]. As a result, computational phenotypes tend to capture only individuals whose gender-related care is both medicalized and well-documented [24,25,30,31]. Those who do not disclose their gender, follow non-normative care pathways, or who avoid or are unable to access gender-affirming care are likely underrepresented. Alpert et al. found that their algorithm "would have identified 0.003% of patients seen at CancerLinQ practices as of October 2019 as transgender," which is a substantial underestimate [29]. This exclusion can in turn distort downstream analyses, for example, by leading to conclusions that trans people are more likely to be white or concentrated in the Northeastern and Western areas of the U.S. [37].

Hines et al. explicitly note that reliance on proxies for medical intervention excludes a "substantial portion of gender-diverse populations" [34] and suggest that incorporating self-reported sex and gender fields can lead to more representative computational phenotypes. However, self-reported data, if available at all, are constrained by the limited response options permitted within most EHR systems. In an analysis from the University of Iowa Hospitals and Clinics, the same authors found that adolescent and young adult populations often report identities that fall outside of these predefined categories (e.g., agender, demiboy, genderqueer, transfeminine) [34]. Kim et al. similarly proposed expanding gender identity fields to include terms such as "nonbinary," "gender fluid," and "unsure/questioning," which frequently appear in free-text entries within patient records [35].

In response to these limitations, researchers often rely on constructed variables that indicate discrepancies between sex and gender fields (e.g., gender identity recorded as 'male' and sex as 'female' [27]) or patterns in prescription data (e.g., male gender marker and estrogens/progestins, estrogen, or progesterone and spironolactone 200 mg [30]). While these efforts aim to better capture gender diversity within the constrained structure of EHR data, they embed assumptions about bodies and medical transition pathways, and fail to account for identities that are nonbinary, fluid, neutral, or evolving over time. Moreover, Foer et al. reported that this approach was particularly inaccurate. At Partners Healthcare in Boston, all patients flagged based solely on discrepancies across sex assigned at birth, legal sex, and gender identity fields were ultimately found to be cisgender upon chart review. [26].

While many studies turned to keywords within clinical notes to address the limitations of structured data, keyword selection generally mirrors prevailing clinical documentation practices, rather than reflecting current or culturally relevant language, particularly for people of color and nonbinary individuals. For example, the computational phenotype developed by Roblin et al. relied on a very narrow set of keywords, including "transgender," "transsexual," "transvestite," "gender," "gender dysphoria," and "gender reassignment" [24]. Although subsequent studies have broadened their keyword sets, it remains impossible to fully capture the diversity of gender terminology, let alone to assume that this diversity is adequately reflected in EHR data. Ehrenfeld et al. speculated that adding more keywords would improve the accuracy of their algorithm (e.g., "nonbinary", "genderqueer"), but pointed out that language is bound to change and that identities will need to be continuously added [25]. Similarly, Xie et al. emphasized that keyword lists will require ongoing revision to remain current with evolving language [28]. In an effort to move beyond keyword-based methods, Hua et al. utilized a deep learning approach to identify gender-diverse individuals without relying on manually selected terms [33]. However, their model struggled with contextual understanding, for example confusing terms like "hysterectomy" and "they/them", and was trained primarily on PubMed and social media posts due to limited access to large-scale EHR data [88]. These issues underscore the limitations of both rule-based and machine learning algorithms when applied to contexts where data on gender are incompletely or inaccurately documented.

Bias in validation

Bias in algorithm design is further compounded by flawed validation practices. Gold-standard labels for gender are most commonly derived from manual review of EHRs by trained annotators, a process referred to as "chart review" (see **Table 2**). This practice arises out of necessity, as self-reported data is rarely fully documented in patient records [11,14,37]. However, using chart review to establish the gold standard assumes that annotators can accurately infer gender from a patient's historical EHR data. While this assumption may be reasonable for well-documented chronic conditions that have traditionally been the focus of computational phenotyping (e.g., asthma, heart failure), it is unlikely to hold true for gender [23,24]. For instance, chart review often only identifies a patient as transgender if there is explicit documentation of gender dysphoria or evidence of medical transition. In one study, Alpert et al. limited their review to records containing relevant diagnosis codes, leading to the misclassification of some transgender patients as "not transgender" [29]. More broadly, the absence of documentation is not neutral [89]. It may reflect erasure, patient mistrust, or systemic failures to solicit or record information.

| Study | Method for Gold Standard | TPR | FPR | PPV | NPV | F1 |
|---|---|---|---|---|---|---|
| Roblin et al. (2016) | Chart Review | | | 0.68 | | |
| Ehrenfeld et al. (2019) | Chart Review | | 0.03 | | | |
| Foer et al. (2020) | Chart Review | 1 | 0.73 | 0.08 | | |
| Chyten-Brennan et al. (2020) | Chart Review | | | 0.84 | | |
| Xie et al. (2020) | Chart Review | 0.97 | 0.05 | 0.95 | 0.97 | 0.96 |
| Alpert et al. (2021) | Chart Review | | | 0.76 | | |
| Guo et al. (2021) | Chart Review | 1 | | 1 | 1 | 1 |
| Wolfe et al. (2021) | Chart Review | | | 0.83 | | |
| Streed et al. (2023) | Self-reported gender and sex assigned at birth | 0.87 | 0.01 | 0.89 | 0.99 | |
| Hua et al. (2023)[†] | Chart Review | 0.97 | 0.10 | 0.99 | 0.75 | 0.98 |
| Hines et al. (2023) | Chart Review | | | 1 | | |
| Kim et al. (2024)[†] | Chart Review | | | | | |

**Table 2**: Accuracy of existing computational phenotypes. Shown is the method used to obtain the gold-standard label and associated performance metrics. Abbreviations: TPR = True Positive Rate, FPR = False Positive Rate, PPV = Positive Predictive Value, NPV = Negative Predictive Value, F1 = F1 score.

† Hua et al. used a machine learning model and also included results for accuracy, AUC, and AUPRC (0.96, 0.86, and 0.99, respectively). Kim et al. used the gold-standard to refine the keywords used in the algorithm and did not perform validation.

These limitations have important implications for the use of computational phenotypes in biomedical research. When algorithms are evaluated against such flawed reference standards, even high accuracy metrics are misleading (see **Table 2**). As an illustrative example, consider a simple analysis aimed at estimating the prevalence of a clinical condition among individuals identified as transgender by a computational phenotype. Even if the algorithm appears perfectly accurate relative to a chart-reviewed gold standard, its utility is compromised if the gold standard itself has low sensitivity. In such cases, many individuals will be excluded from both the gold standard and the algorithm's outputs, leading to biased estimation of the disease prevalence and a distorted understanding of the condition's impact on trans populations. This example highlights how limiting biomedical research to individuals who are legible to algorithms can perpetuate incomplete or skewed representations of transgender health and, in doing so, obscure the very populations that computational phenotypes intend to make visible.

Unfortunately, this concern is not hypothetical. In a related study, Manfredi et al. used insurance claims data to examine cancer outcomes among transgender women, reporting a lower prevalence of prostate cancer among those receiving gender-affirming hormone therapy, but a positive correlation between hormone use and aggressive disease [90]. In response, Hamnvik et al. and Berner et al. raised methodological concerns in two separate letters to the editor [38,39]. Foremost among them was the use of diagnosis codes to identify transgender women. These codes are known to have low sensitivity and may also capture individuals undergoing treatment for other conditions, such as orchiectomy or prostate cancer, resulting in substantial misclassification. The study's finding that only 31.5% of transgender women had records of hormone therapy, compared to 71% in similar datasets, further underscores this concern [90]. As both sets of authors argue, such misclassification not only compromises the validity of the study's findings, but also risks reinforcing harmful narratives that could jeopardize access to hormone therapy for transgender women.

## Potential for misuse

*"[S]urveillance is a central practice through which the category of transgender is produced, regulated, and contested." [91]*

In addition to methodological challenges, computational phenotyping raises significant ethical concerns, particularly regarding the potential misuse of algorithmic outputs. Although not consistently addressed across studies, Chyten-Brennan et al. highlight the risk of identifying transgender and nonbinary patients in environments of pervasive discrimination [27]. Reflecting these concerns, Hua et al. excluded individuals who chose not to disclose information in sex and gender fields when developing their phenotyping algorithm [33]. More broadly, Hines et al. point out the dangers of recording gender identity in EHRs, citing risks such as bias, discrimination, and mistreatment by healthcare providers [34]. These risks are especially pertinent for children and adolescents, whose medical information is accessible to parents or guardians [73].

In related work, Alpert et al. examine the principle of beneficence in transgender health research using insurance claims data [37]. They point out that even de-identified datasets carry a "theoretical, but plausible" risk of reidentification, particularly within small populations [92,93]. Critically, the authors also highlight that computational phenotyping can involve identifying individuals receiving gender-affirming care without an explicit diagnosis, such as those with a code for an unspecified endocrine disorder or relevant keywords in clinical notes [37]. In the current political climate marked by the criminalization of gender-affirming care [94], attempts to access transgender patients' health records [95], and systemic efforts to erase trans identities, computational phenotyping can become a tool of surveillance that amplifies discrimination, misclassification, and inflicts harm [96,97].

Similar warnings have been made in the context of automated gender recognition tools, such as those used in airport body scanners [91]. Scholars of technology and ethics have broadly critiqued such "processes of technologization and rationalization that frame bodies, identities, and groups as outside of historical frameworks and experiences of racial and other forms of difference" [98]. Likewise, computational phenotyping, when abstracted away from its broader sociopolitical context, risks reinforcing the structural conditions it purports to address. At this moment, when the stakes are not merely theoretical, interdisciplinary researchers and scholars must seriously evaluate potential benefits of algorithmic development against the substantial and potentially life-threatening risks to already vulnerable populations [56,99]. This is particularly evident in ongoing debates over algorithmic race classification, which offer several important lessons for computational phenotyping.

## Lessons from algorithmic race classification

Within healthcare and biomedical research, algorithmic race classification is the automated and predictive assignment of race from proxies like demographics and clinical data, often without self-identification or consent [100,101]. For example, Gichoya et al. found that deep learning models can infer self-reported race from medical imaging data alone, even when images are cropped, corrupted, or noised, and with performance generalizing across imaging modalities and healthcare settings [101]. This finding poses an enormous risk, as such models are a direct vessel for the reproduction and exacerbation of race-based disparities that exist within healthcare. The danger is further compounded by the fact that human oversight is of limited use to recognize and mitigate these race-based disparities as clinicians cannot accurately identify racial identity from medical images themselves. As the authors warn, "if an AI model relies on its ability to detect racial identity to make medical decisions, but in doing so produced race-specific errors, clinical radiologists (who do not typically have access to racial demographic information) would not be able to tell, potentially leading to errors in health-care decision processes" [101].

Critical scholars caution against algorithmic race classification altogether on the premise that the practice risks reinscribing race as a biological concept, which is an outdated pseudoscientific claim that has historically been used to rationalize slavery, eugenics, and other social inequalities [77,102–105]. Other researchers point out that even self-identified or self-reported race does not necessarily align with biological traits, further highlighting the conceptual contradictions underlying algorithmic race classification [106]. While Gichoya et al. acknowledge that race is a social rather than a biological construct, and that more genetic variation exists within racial groups than between them, they also maintain that self-reported race remains a strong proxy for racial identity [101]. This claim risks reifying race as a biological construct or causal factor rather than a fluid social construct. Many critical scholars instead situate algorithmic race classification in a political context, shaped by colonial logics of surveillance, governance, and control [103,107,108]. From this perspective, race-classifying algorithms do not merely reflect social categories, they reproduce and automate racial hierarchies under the guise of neutrality and objectivity, perpetuating racist science and deepening the marginalization of racialized communities [77,102].

Similarly, computational phenotyping rests on the premise that gender can be predicted from proxies rather than attempting to capture gender as social, fluid, and self-determined [23]. In both cases, identity is rendered visible through logics of surveillance, which omits the opportunity to disclose identity through individual agency. Applied to gender, this surveillance logic further extends what Ruha Benjamin calls the "New Jim Code," which refers to the coded and automated reinforcement of inequities through technical systems that are deeply embedded with racialized and gendered assumptions, but are made to appear neutral or objective [102]. As Os Keyes argues in *The Misgendering Machines*, algorithmic systems that attempt to classify gender often make ontological claims about what gender is, reducing it to a binary and ignoring its fluid and socially constructed nature, thereby excluding and harming those who identify as non-binary [77]. These systems also presume that gender is physiologically rooted, essentializing the body as the source of truth, further harming and discriminating against those who identify as trans. These issues strongly parallel the problems with algorithmic race classification, where harm exists not only in the errors made by the algorithms, but also with proxy-based and data-driven construction of identity, which reproduces categories that historically and contemptuously pathologize racialized and gendered individuals in ways that are bound by outdated and harmful colonial logics. Rather than reform these practices through community-based inclusivity or more representational data sets, many critical scholars across Black studies, data justice studies, and trans studies urge us to challenge the legitimacy and necessity of algorithmic classification itself, especially considering the historical and political contexts in which it is situated and the potential harms that are at risk of being produced [77,102,103,107,108].

## Conclusions

 *"When approaching any new source of knowledge…it's essential to ask questions about the social, cultural, historical, institutional, and material conditions under which that knowledge was produced…"* [56]

While computational phenotyping of gender is increasingly used in EHR-based biomedical research, it raises significant methodological and ethical concerns that challenge the validity, and ultimately the utility, of this practice. Phenotyping attempts to infer gender through clinical, biological, or administrative proxies, which is methodologically flawed and conceptually problematic. It also risks perpetuating a history of using gender in ways that have contributed to scientific misrepresentation and social injustice, making even well-intentioned studies susceptible to causing harm. We close by outlining existing recommendations for biomedical researchers and identifying priorities for future research.

### Existing recommendations

In related work examining the ethics of identifying and researching transgender and gender-diverse individuals from insurance claims data, Alpert et al. apply the framework of epistemic justice to propose six recommendations for minimizing harm and maximizing benefits for transgender individuals and communities (see **Table 3**) [37]. Generally, the reviewed studies did not follow all of these principles. For example, only one study in our review included a positionality statement identifying the lead author as a cisgender male [35]. Similarly, Cato et al. invoke the Belmont Report's principles of respect for persons, beneficence, and justice to highlight a range of ethical issues in EHR phenotyping more broadly [109]. These issues include patient consent for secondary data use, the balance of harms and benefits in research based on phenotyping, and the influence of clinician bias, whether conscious or unconscious, on study design and findings. The authors advocate for greater community consultation, transparency in data use, privacy-preserving approaches, and dynamic or portable consent models that return more control to patients. Comparable calls have emerged outside the EHR phenotyping literature [110]. For example, building on decades of scholarship on how science can serve marginalized populations, Kennis et al. propose four concrete actions for researchers: establishing advisory boards with transgender representation, assembling multidisciplinary teams, prioritizing life-saving research, and restructuring the ethical approval process [111]. While these recommendations are valuable and actionable, we build on them by taking a complementary, though more critical, stance informed by recent scholarship conceptualizing how gender is understood and operationalized in scientific research.

| |
|---|
| 1. Explicitly describe the categories that are utilized (e.g., people with cervixes as evidenced by procedure codes). |
| 2. Explicitly acknowledge data limitations and dangers to transgender communities. |
| 3. Use reflexivity by which researchers state their positionality and biases to contextualize their work. |
| 4. Prevent identifiability of transgender individuals. |
| 5. Transgender researchers—especially those with multiple marginalized identities—lead or co-lead research, guide analyses, and interrogate the work's ethics and utility. |
| 6. Supplement claims-based research conducted on transgender people with community-based participatory research conducted by and alongside transgender people. |

**Table 3**: Suggestions from Alpert et al. [37] to minimize harm and maximize benefit to transgender individuals and communities when using insurance claims data for biomedical research.

## Priorities for future research

In a 2024 *Nature* special collection exploring the risks and challenges of integrating of sex and gender into research, Ashley et al. highlight the inadequacy of current terminology for gender-based research, arguing that it lacks both the pragmatism required for scientific inquiry and the flexibility needed to reflect the diversity of human experience [75]. To address this gap, the authors introduce the concept of *gender modality* [112], defined as the relationship between a person's gender identity and the gender assigned at birth. Much like the concept of sexual orientation which has moved us away from a gay/straight binary, this framework includes familiar categories like cisgender and transgender, while also capturing a broader range of experiences (see **Table 4**). Gender modality is a concept that is already in use by transgender communities, clinicians, and policymakers, and has been applied by Statistics Canada, Planned Parenthood, and the Supreme Court of Canada [75]. Ashley et al. argue that this shift in terminology can improve scientific inquiry in three ways: (1) expanding how gender is categorized and captured in data, (2) refining research questions and interpretations, and (3) forcing greater clarity on what investigators are actually measuring. While no single framework can resolve all challenges, establishing more nuanced language is an essential step toward biomedical research that embraces, rather than simplifies, the complexity of gender.

| Modality | Definition |
|---|---|
| Agender | People who do not identify with any gender. |
| Cisgender | People whose gender identity corresponds to the gender they were assigned at birth. |
| Closeted trans people | Individuals whose gender identity does not correspond to the gender they were assigned at birth, but who do not share their gender identity publicly. |
| Culture-specific identities | Individuals can have identities, such as Two-Spirit identities in North American Indigenous communities and hijra on the Indian subcontinent, that might not align with Western concepts of gender and sexuality. People with these identities might not consider themselves cis or trans because of the Western philosophies that underpin these terms. |
| Detrans/retrans | People who have ceased, shifted or reversed their gender transition. |
| Gender questioning | People who are unsure of their gender identity and are in the process of working it out. |
| Intersex | People who were born or who endogenously developed sexual traits that differ from typical expectations of female and male bodies. Some intersex people do not consider themselves to be cis or trans. |

| People with dissociative identity disorder whose alters have distinct gender identities | People with this condition, also known as plural people, can have several identities, known as alters or headmates, that have distinct gender identities. These alters can have different gender modalities. |
|---|---|
| Raised in a gender-neutral manner | People who were raised without being referred to as a boy/he or girl/she until they were old enough to express their gender identity. |
| Transgender | People whose gender identity does not correspond to the gender they were assigned at birth. |

**Table 4**: A non-exhaustive list of gender modalities provided in Ashley et al. [75].

The lack of adequate terminology underscores a more fundamental problem: gender has yet to be fully conceptualized within scientific practice, and it is unclear whether a complete conceptualization is currently possible or desirable. As noted by Restar et al. in the context of epidemiological studies, "[M]easuring gender and sex has no gold standard, perhaps since these variables depend on time and context. Pragmatically, research questions, aims, scope, study design, methods, and capacity to collect and analyze data should all influence how to measure gender and sex. That is, as there is no single best practice, investigators must decide which dimensions of sex and gender are relevant to their research questions." Numerous scholars have also challenged the validity of gender variables in science more broadly [77,102,103], arguing that existing categories are embedded in histories of pseudoscience and structural injustice. Therefore, measurement can never be "fixed" within existing systems, but instead must be reimagined or abandoned entirely. Within EHR-based research, rather than attempting to reverse-engineer gender with computational phenotypes from distorted and incomplete data shaped by a legacy of structural oppression, we argue that researchers should shift their focus toward developing a just and conceptually sound foundation for gender-based research.

Practically, this means creating and using measurement tools that accommodate fluidity, center lived experience rather than biological proxies, and allow for individualized data collection without defaulting to gender assignment. This echoes efforts like Kronk et al.'s framework for transgender data collection in EHR systems [14], as well as rethinking gender measurement more broadly through the lenses of data justice and intersectional feminist and queer theory [56,113]. With access to larger and larger health data sets, researchers and scholars must collectively welcome and critically engage with questions that appear deceptively simple, such as: What exactly are we trying to measure? Can it be measured? How has it been measured before, and who might be harmed by these approaches? Without grounding scientific work in ethical frameworks, sociopolitical context, and epistemic reflexivity, we risk perpetuating the very structures of marginalization that we seek to challenge. [79]

# Declarations

## Ethics approval and consent to participate

Not applicable.

## Consent for publication

Not applicable.

## Availability of data and materials

No datasets were used or analyzed for this study.

## Competing interests

The authors declare that they have no competing interests.

## Funding

This study was supported by a Critical Investigation of Data Science Grant from the University of Toronto Data Science Institute awarded to J.G.

## Author's Contributions

JG and HT conceived the study. JG designed the study and provided supervision to LD and DC. JG, HT, LD, and VF drafted the manuscript. LD and DC performed the full-text review. All authors revised and provided valuable feedback on the manuscript.

## Acknowledgments

Not applicable.

# Abbreviations

EHR = Electronic health record
Trans, TG = Transgender
MTF = Male to Female
FTM = Female to Male
TGNB = Transgender and Nonbinary
TGNC = Transgender and Gender-Nonconforming
TM = Transmasculine
TF = Transfeminine
TGD = Transgender and Gender Diverse
TGM = Transgender Men

TGW = Transgender Women
GE = Gender Expansive.